\documentstyle[aps,pra,multicol]{revtex}
\begin{document}

\draft
\title{Towards Strong-Coupling Generalization
of the Bogoliubov Model}
\author{A. Yu. Cherny$^{1}$ and A. A. Shanenko$^{2}$ }
\address{$^{1}$Frank Laboratory of Neutron Physics,
Joint Institute for Nuclear Research,
141980, Dubna, Moscow region, Russia}
\address{$^{2}$Bogoliubov Laboratory of Theoretical Physics,
Joint Institute for Nuclear Research,
141980, Dubna, Moscow region, Russia}
\date{June 28, 1998}

\maketitle

\begin{abstract}
The well-known results concerning a dilute Bose gas with the
short-range repulsive interaction should be reconsidered due to a
thermodynamic inconsistency of the method being basic to much of the
present understanding of this subject and nonrelevant behaviour of
the pair distribution function at small boson separations.  The aim
of our paper is to propose a new way of treating the dilute Bose gas
with an arbitrary strong interaction.  Using the reduced density
matrix of the second order and a variational procedure, this way
allows us to escape the inconsistency mentioned and operate with
singular potentials like the Lennard-Jones one. All the
consideration concerns the zero temperature.
\end{abstract}
%\pacs{PACS numbers: 05.30.Jp, 05.30.-d, 03.75.Fi}

\begin{multicols}{2}
%-----------------------------------------------------------------
\section{Introduction and basic equations}
\label{1sec}

It is well-known that to investigate a dilute Bose gas of particles
with an arbitrary strong repulsion (the strong-coupling regime),
one should go beyond the Bogoliubov approach~\cite{Bog1} (the
weak-coupling case) and treat the short-range boson correlations
in a more accurate way. An ordinary manner of doing so is the use of
the Bogoliubov model with the ``dressed", or effective, interaction
potential containing ``information" on the short-range boson
correlations (see Ref. \cite{Lee}). Below it is demonstrated that
this manner leads to a loss of the thermodynamic consistency. To
overcome this trouble, we propose a new way of investigating the
strong-coupling regime which concerns the reduced density matrix of
the second order~(the 2-matrix) and is based on the variational
method.

The 2-matrix for the many-body system of spinless bosons can be
represented as~\cite{Bog2}:
\begin{equation}
\rho_2({\bf r}_1^{\prime},{\bf r}_2^{\prime};{\bf r}_1,{\bf r}_2)
= {F_2({\bf r}_1,{\bf r}_2;{\bf r}_1^{\prime},{\bf r}_2^{\prime})
\over N(N-1)},
\label{1}
\end{equation}
where the pair correlation function is given by
\begin{equation}
F_2({\bf r}_1,{\bf r}_2;{\bf r}_1^{\prime},{\bf r}_2^{\prime})=
\langle \psi^{\dagger}({\bf r}_1) \psi^{\dagger}({\bf r}_2)
    \psi ({\bf r}_2^{\prime})\psi ({\bf r}_1^{\prime})\rangle.
\label{2}
\end{equation}
Here $\psi ({\bf r})$ and $\psi^{\dagger}({\bf r})$ denote the boson
field operators. Recently it has been found~\cite{Chern1,Chern2}
that for the uniform system with a small depletion of the
zero-momentum state the correlation function (\ref{2}) can be
written in the thermodynamic limit as follows:
\begin{eqnarray}
F_2({\bf r}_1,{\bf r}_2;{\bf r}_1^{\prime},{\bf r}_2^{\prime})&=&
n_0^2\,\varphi^*(r)\,\varphi(r^{\prime})\nonumber\\
&&+2n_0\int \frac{d^3q}{(2\pi)^3}\,
       n_q\,\varphi_{{\bf q}/2}^*({\bf r})
            \varphi_{{\bf q}/2}({\bf r}^{\prime})\nonumber\\
&&\times\exp\{ i {\bf q}({\bf R}^{\prime}-{\bf R})\},
\label{3}
\end{eqnarray}
where ${\bf r}={\bf r}_1 - {\bf r}_2, \;{\bf R}=({\bf r}_1
+{\bf r}_2)/2$
and similar relations take place for ${\bf r}^{\prime}$ and ${\bf
R}^{\prime}$, respectively. In Eq.~(\ref{3}) $n_0=N_0/V$ is the
density of the particles in the zero-momentum state, $n_q=\langle
a_{\bf q}^{\dagger}a_{\bf q}\rangle$ stands for the distribution of
the noncondensed bosons over momenta. Besides, $\varphi(r)$ is the
wave function of a pair of particles being both condensed. In turn,
$\varphi_{{\bf q}/2}({\bf r})$ denotes the wave function of the
relative motion in a pair of bosons with the total momentum
$\hbar{\bf q}$, this pair including one condensed and one
noncondensed particles.
So, Eq.~(\ref{3}) takes into account the condensate-condensate and
supracondensate-condensate pair states and is related to the
situation of a small depletion of the zero-momentum one-boson
state. For the wave functions $\varphi(r)$ and $\varphi_{{\bf p}}
({\bf r})$ we have
\begin{eqnarray}
\varphi(r)\!=\!1+\psi(r),\ \varphi_{{\bf p}}\!({\bf r})\!=\!
\sqrt{2}\cos({\bf p}{\bf r})+\psi_{{\bf p}}({\bf r})\;(p\not=0)
\label{4}
\end{eqnarray}
with the boundary conditions $ \psi(r) \to 0$ and
$\psi_{{\bf p}}({\bf r}) \to 0$ for $r \to \infty.$
The functions $\psi(r)$ and $\psi_{{\bf p}}({\bf r})$ can
explicitly be expressed in terms of the Bose operators $a_{{\bf
p}}^{\dagger}$ and $a_{{\bf p}}$ \cite{Chern1}:
\begin{eqnarray}
\widetilde{\psi}(k)
&=&\int\psi(r)\exp(-i{\bf k}{\bf r})\,d^3r=
  \langle a_{{\bf k}}\,a_{-{\bf k}}\rangle/n_0,
\label{6}\\
\widetilde{\psi}_{\bf p}({\bf k})
&=&\int\psi_{{\bf p}}({\bf r})\exp(-i{\bf k}{\bf r})\,d^3r \nonumber\\
&&=\sqrt{\frac{V}{2 n_0}}
\frac{\langle a^{\dagger}_{2{\bf p}}
a_{{\bf p}+{\bf k}} a_{{\bf p}-{\bf k}}\rangle}{n_{2p}}.
\label{44}
\end{eqnarray}
Having in our disposal the distribution function $n_{k}$ and the set
of the pair wave functions $\varphi(r)$ and $\varphi_{{\bf p}} ({\bf
r})$, we are able to calculate the main thermodynamic quantities of
the system of interest. In particular, the mean energy per particle
is expressed in terms of $n_k$ and $g(r)$ via the well-known formula
\begin{eqnarray}
\varepsilon=\int \frac{d^3k}{(2\pi)^3} T_k \frac{n_k}{n}+
\frac{n}{2}\int g(r) \Phi(r) d^3r,
\label{7}
\end{eqnarray}
where $T_k=\hbar^2 k^2/2m$ is the one-particle kinetic energy,
$n=N/V$ stands for the boson density and the relation
\begin{equation}
g(r)=F_2({\bf r}_1,{\bf r}_2;{\bf r}_1,{\bf r}_2)/n^2.
\label{8}
\end{equation}
is valid for the pair distribution function $g(r)$.

%-----------------------------------------------------------------
\section{The Bogoliubov model}
\label{2sec}

The starting point of our investigation is the weak-coupling regime
which implies weak spatial correlations of particles and, thus, is
characterized by the set of the inequalities
\begin{equation}
|\psi(r)| \ll 1, \;\quad |\psi_{{\bf p}}({\bf r})| \ll 1\;.
\label{9}
\end{equation}
Specifically, the Bogoliubov model corresponds to the
choice~\cite{Chern1,Chern2}
\begin{equation}
|\psi(r)| \ll 1, \;\quad \psi_{{\bf p}}({\bf r}) = 0\;.
\label{10}
\end{equation}
Besides, owing to a small depletion of the Bose condensate
$(n-n_0)/n$ we have for the one-particle density matrix
$F_{1}(r)=\langle\psi^{\dagger}({\bf r}_{1})\psi({\bf
r}_{2})\rangle$:
$$
\left|\frac{F_{1}(r)}{n}\right|=
\left|\int\frac{d^3k}{(2\pi)^3}\frac{n_k}{n}
\exp(i{\bf k}{\bf r})\right|\leq \frac{n-n_{0}}{n}\ll 1.
$$
So, investigating the Bose gas within the Bogoliubov scheme, we have
two small quantities: $\psi(r)$ and $F_{1}(r)/n$.  This enables us
to write Eq.~(\ref{8}) with the help of (\ref{3}) as follows:
\begin{equation}
g(r)=1+2\psi(r)+\frac{2}{n}
\int\frac{d^3k}{(2\pi)^3}n_k\exp(i{\bf k}{\bf r}),
\label{11}
\end{equation}
where we restricted ourselves to the terms linear in $\psi(r)$ and
$F_{1}(r)/n$ and put $\psi^*(r)=\psi(r)$ because the pair wave
functions can be chosen as real quantities.  Equations for
$\widetilde{\psi}(k)$ and $n_k$ can be found varying the mean energy
(\ref{7}) with Eq. (\ref{11}) taken into account. However, previously
one should realize an important point, namely: $n_k$ and
$\widetilde{\psi}(k)$ can not be independent variables.  Indeed,
when there is no interaction between particles, there are no spatial
particle correlations either. So, $\widetilde{\psi}(k)=0$ and, since
the zero-temperature case is considered, all the bosons are
condensed, $n_k=0$. While ``switching on" the interaction results
in appearing the spatial correlations and condensate depletion:
$\widetilde{\psi}(k)\not=0$ together with $n_k\not=0$.  In the
framework of the Bogoliubov scheme $\widetilde{\psi}(k)$ is related
to $n_k$ by the expression
\begin{equation}
n_k (n_k+1)=n_0^2\widetilde{\psi}^2(k).
\label{12}
\end{equation}
Indeed, the canonical Bogoliubov transformation~\cite{Bog1}
implies that
\begin{equation}
a_{{\bf k}}=u_{k}\alpha_{{\bf k}}+v_{k}
\alpha^{\dagger}_{-{\bf k}},\quad
a^{\dagger}_{{\bf k}}=u_{k}\alpha^{\dagger}_{{\bf k}}+
v_{k}\alpha_{-{\bf k}},
\label{12a}
\end{equation}
where
\begin{equation}
u_{k}^2-v_{k}^2=1.
\label{13}
\end{equation}
At zero temperature $\langle\alpha^{\dagger}_{{\bf k}}
\alpha_{{\bf k}} \rangle=0$ and, using Eqs.~(\ref{6}) and (\ref{12a})
we arrive at
\begin{equation}
n_k=v_{k}^2, \quad
      \widetilde{\psi}(k)=u_{k}v_{k}/n_0.
\label{14}
\end{equation}
With Eqs.~(\ref{13}) and (\ref{14}) one can readily obtain
Eq. (\ref{12}).

Now, let us show that all the results on the thermodynamics of a
weak-coupling Bose gas can be derived for the Bogoliubov scheme with
variation of the mean energy (\ref{7}) under the conditions
(\ref{11}) and (\ref{12}). Inserting Eq. (\ref{11}) into Eq.
(\ref{7}) and, then, varying the obtained expression, we arrive at
\begin{equation}
\delta\varepsilon=\int \frac{d^3k}{(2\pi)^3} \left\{\Bigl(T_k+
n\widetilde\Phi(k)\Bigr)\frac{\delta n_k}{n} + n\widetilde\Phi(k)
\delta\widetilde\psi(k)\right\}.
\label{15}
\end{equation}
Relation (\ref{12}) connecting $\widetilde\psi(k)$ with $n_k$
results in
\begin{equation}
\delta\widetilde{\psi}(k)=\frac{(2n_k+1)\delta n_k}{2n_0^2
\widetilde{\psi}(k)}+
\frac{\widetilde{\psi}(k)}{n_0}\int\frac{d^3q}{(2\pi)^3}\delta n_q,
\label{16}
\end{equation}
where the equality
\begin{equation}
n=n_0+\int\frac{d^3k}{(2\pi)^3}\;n_k
\label{17}
\end{equation}
is taken into consideration.
Setting $\delta\varepsilon=0$ and using Eqs.~(\ref{15}) and
(\ref{16}), we derive the following expression:
\begin{eqnarray}
-2\,T_k\widetilde{\psi}(k)
&=&\frac{n^2}{n^2_0} \widetilde{\Phi}(k)(1+2n_k)
+2n\,\widetilde{\psi}(k)\nonumber\\
&&\times\left(\widetilde{\Phi}(k)+\frac{n}{n_0}
\int\frac{d^3q}{(2\pi)^3}\widetilde{\Phi}(q)\widetilde{\psi}(q)\right).
\label{18}
\end{eqnarray}
Here one should realize that Eq.~(\ref{18}) is able to yield results
being accurate only to the leading order in $(n-n_0)/n$ because the
used expression for $g(r)$ given by Eq. (\ref{11}) is valid to the
next-to-leading order~\cite{note4}. So, Eq.~(\ref{18}) should be
rewritten as
\begin{equation}
-2\,T_k \widetilde{\psi}(k)=
\widetilde{\Phi}(k)(1+2n_k)+2n\,\widetilde{\psi}(k)\Phi(k).
\label{19}
\end{equation}
Equation (\ref{19}) is an equation of the Bethe-Goldstone type or, in
other words, the in-medium Schr\"odinger equation for the pair wave
function. As $2\widetilde{\Phi}(k)(n_k+n \widetilde{\psi}(k))$ is
the product of the Fourier transforms of $\Phi(r)$ and $n(g(r)-1)$,
we can rewrite Eq.~(\ref{19}) in the more customary form
\begin{equation}
\frac{\hbar^2}{m}\nabla^2\varphi(r)=\Phi(r)
+n\int\Phi(|{\bf r}-{\bf y}|)\Bigl( g(y)-1\Bigr)d^3y.
\label{20}
\end{equation}
The structure of Eq.~(\ref{20}) is discussed in the
papers~\cite{Chern2,Shan1}. Here we only remark that the right-hand
side (r.h.s.) of Eq. (\ref{20}) is the in-medium potential of the
boson-boson interaction in the weak-coupling approximation. The
system of equations (\ref{12}) and (\ref{19}) can easily be solved,
which leads to the familiar results~\cite{Bog1}:
\vspace*{-2mm}
\begin{eqnarray}
&&n_k=\frac{1}{2}\left(\frac{T_k+n\widetilde{\Phi}(k)}{\sqrt{T_k^2+
2nT_k\widetilde{\Phi}(k)}}-1\right),\nonumber\\[-1mm]
&&\widetilde{\psi}(k)=-\frac{
\widetilde{\Phi}(k)}{2\sqrt{T_k^2+2nT_k\widetilde{\Phi}(k)}}\,.
\label{21}
\end{eqnarray}

%----------------------------------------------------------
\section{A dilute Bose gas within the Bogoliubov model}
\label{3sec}

As it was mentioned, the aim of our paper is investigation of the
case of a dilute Bose gas with an arbitrary strong repulsion between
bosons. So, considering a dilute Bose gas in the weak-coupling
approximation can be a good exercise providing us with useful
information. Let us investigate the thermodynamics of a dilute
Bose gas within the Bogoliubov model. With Eqs.~(\ref{7}),
(\ref{11}) and (\ref{21}) we derive
\begin{eqnarray}
\varepsilon&=&\frac{n}{2}\widetilde{\Phi}(0)
+\frac{1}{2n}\int\frac{d^3k}{(2\pi)^3}\nonumber\\
&&\times\left(\sqrt{T_k^2+2nT_k\widetilde{\Phi}(k)}-T_k -
                 n\widetilde{\Phi}(k)\right).
\label{21a}
\end{eqnarray}
The well-known argument of Landau (see the footnote in
Ref.~\cite{Bog1} and discussion in Ref.~\cite{Lee}) testifies that
the properties of dilute quantum gases are ruled by the scattering
length. Within the Bogoliubov model this length is usually assumed
to be equal to $m\widetilde{\Phi}(0)/4\pi \hbar^2$.  If so, when
expanding $\varepsilon$ in powers of the boson density $n$, one
could replace $\widetilde{\Phi}(k)$ by $\widetilde{\Phi}(0)$ in Eq.
(\ref{21a}), introducing the low-momentum approximation. However,
this leads to a divergency because at large $k$ the integrand
behaves as $-n^2\widetilde{\Phi}^{2}(k)/2T_k$. To properly calculate
the integral in Eq.~(\ref{21a}), we should rewrite Eq. (\ref{21a})
in the following form:
\begin{eqnarray}
\varepsilon=\frac{n}{2}\left(\widetilde{\Phi}(0)-
    \int\frac{d^3k}{(2\pi)^3}
         \frac{\widetilde{\Phi}^2(k)}{2T_k}\right)+I,
\label{21b}
\end{eqnarray}
where
\begin{eqnarray}
I&=&\frac{1}{2n}\int\limits_{0}^{\infty}dk\frac{4\pi k^2}{(2\pi)^3}
\nonumber \\
&&\times\Bigl(\sqrt{T_k^2+2nT_k\widetilde{\Phi}(k)}
-T_k - n\widetilde{\Phi}(k)+
         \frac{n^2\widetilde{\Phi}^2(k)}{2T_k}\Bigr).
\nonumber
\end{eqnarray}
Now, substituting $k=(2 m n y)^{1/2}/\hbar$ in the integral, we
obtain the expression
\begin{eqnarray}
I&=&\frac{\sqrt{2}}{4\pi^2}\left(\frac{m n}{\hbar^2}\right)^{3/2}
\int\limits_{0}^{\infty} dy\biggl\{y\sqrt{y+
   2\widetilde{\Phi}(\sqrt{2m n y}/\hbar)}\nonumber\\
&&-y^{3/2}
-\widetilde{\Phi}(\sqrt{2m n y}/\hbar)\,y^{1/2}+
        \frac{\widetilde{\Phi}^2(\sqrt{2m n y}/\hbar)}{2\sqrt{y}}
\biggr\}
\nonumber
\end{eqnarray}
which at sufficiently small $n$ may be rewritten as
\begin{eqnarray}
I&=&\frac{\sqrt{2}}{4\pi^2} \left(\frac{m n}{\hbar^2}\right)^{3/2}
\nonumber \\
&&\times\int\limits_{0}^{\infty} dy \Bigl\{
          y\sqrt{y+2\widetilde{\Phi}(0)}
-y^{3/2}-\widetilde{\Phi}(0)y^{1/2}
+\frac{\widetilde{\Phi}^2(0)}{2\sqrt{y}}\Bigr\}.
\nonumber
\end{eqnarray}
The derived integral is readily calculated. The result is given
by
\begin{equation}
I=\frac{8}{15\pi^2} \left(\frac{m n}{\hbar^2}\right)^{3/2}
\widetilde{\Phi}^{5/2}(0).
\label{21e}
\end{equation}
In turn, the first term in the r.h.s. of Eq.~(\ref{21b})
can be represented as
\begin{equation}
\frac{n}{2}\left(\widetilde{\Phi}(0)-
    \int\frac{d^3k}{(2\pi)^3}
         \frac{\widetilde{\Phi}^2(k)}{2T_k}\right)=
\frac{n}{2} \int \varphi^{(0)}(r) \Phi(r) d^3r,
\label{21f}
\end{equation}
where $\varphi^{(0)}$ is the solution of Eq.~(\ref{20}) in the
limit $n \to 0$. This is nothing else but the Schr\"odinger
equation in the Born approximation. According to the relations
(\ref{21e}) and (\ref{21f}) we has to conclude that the scattering
length in the case of interest is expressed in the form
\begin{equation}
a_B=\frac{m}{4 \pi \hbar^2} \int \varphi^{(0)}(r) \Phi(r) d^3r.
\label{21g}
\end{equation}
One can easily be convinced that $\widetilde{\Phi}(0)$ can not
be represented only in terms of $a_B$ and, hence, the dependence
on the shape of the interaction potential appears in the series
expansion for the mean energy in the first correction to the term
$2\pi \hbar^2 a_B n/m$. To rewrite our result for $\varepsilon$
in a graphic form, we introduce one more characteristic length
$b>0$ which obeys the relation
\begin{equation}
b=-\frac{m}{4\pi \hbar^2}\int \psi^{(0)}(r) \Phi(r) d^3r=
\frac{m}{4\pi \hbar^2}\int \frac{d^3 k}{(2\pi)^3}
                                  \frac{\Phi^2(k)}{2T_k},
\label{21h}
\end{equation}
where $\psi^{(0)}(r)=\varphi^{(0)}(r)-1$. Further, with the help
of Eqs.~(\ref{21b})-(\ref{21h}), we arrive at
\begin{equation}
\varepsilon=\frac{2\pi \hbar^2 a_B n}{m}
\left\{1+\frac{128}{15\sqrt{\pi}}\sqrt{n a_B^3}\left(1+
\frac{5b}{2 a_B}\right)+\cdots \right\},
\label{21i}
\end{equation}
here the condition $b \ll a_B$ is of use.
It is not difficult to see that the expression (\ref{21h}) taken
with negative sign is the next correction to the scattering length
calculated within the Born approximation. As to the Eq.~(\ref{21g}),
it is related to the next-to-Born approximation. Stress that in the
Bogoliubov model the energy term $n\widetilde{\Phi}(0)/2$ is treated
as the major one~\cite{Bog1}, which implies that the condition
$b \ll a_B$ is fulfilled. This qualitative criterion can be written
as
\begin{equation}
\widetilde{\Phi}(0) \gg \int \frac{d^3k}{(2\pi)^3}
      \frac{\widetilde{\Phi}^2(k)}{2T_k}.
\label{21j}
\end{equation}
Beyond this inequality the model may be thermodynamically unstable.
In particular, the opposite case
$$
\widetilde{\Phi}(0) < \int \frac{d^3k}{(2\pi)^3}
      \frac{\widetilde{\Phi}^2(k)}{2T_k}
$$
leads to the negative scattering length (\ref{21g}) which at
sufficiently low densities results in
$-\partial^{2}E/\partial V^{2}=\partial p/\partial V > 0$.

Thus, investigated within the Bogoliubov model, the thermodynamics
of a dilute Bose gas is ruled by the scattering length only in
the zero-density limit. While the next-to-leading term in
the series expansion given by Eq. (\ref{21i}) depends on the shape of
the interaction, which is expressed in appearance of the additional
characteristic length $b$. This conclusion differs from the
results of papers~\cite{Lee} according to which the series expansion
for $\varepsilon$ taken to the same order as that of Eq. (\ref{21i}),
is fully determined by the scattering length. To clarify the
situation concerning this difference, we should go to the
strong-coupling regime.

%--------------------------------------------------------------
\section{The strong-coupling regime}
\label{4sec}

Now, after the detailed investigations of the Bogoliubov model
within the scheme proposed, we are able to demonstrate that the
investigation of the strong-coupling case based on the Bogoliubov
model with the effective boson-boson interaction, results in a loss
of the thermodynamic consistency. Indeed, as it was shown in the
previous section, any calculating scheme using the basic relations
of the Bogoliubov model (\ref{11}), (\ref{12}) conclusively leads to
Eqs.~(\ref{19})-(\ref{21}) provided this scheme does yield the
minimum of the mean energy.  In this case Eqs.~(\ref{19})-(\ref{21})
certainly includes the quantity $\Phi(r)$ which is the ``bare"
interaction potential appearing in Eq. (\ref{7}). The use of the
Bogoliubov model with the effective interaction potential
substituted for $\Phi(r)$ can in no way disturb the relations given
by Eqs. (\ref{11}) and (\ref{12}). And Eq.~(\ref{7}) is the same in
both the weak- and strong-coupling regimes. Thus, any attempts of
replacing $\Phi(r)$ by the effective ``dressed" potential without
modifications of Eqs. (\ref{11}) and (\ref{12}) results in a
calculating procedure which does not really provide the minimum of
the mean energy. It is nothing else but a loss of the thermodynamic
consistency. We remark that we do not mean, of course, that the
t-matrix approach or the pseudopotential method can not be applied
in the quantum scattering problem. It is only stated that the usual
way of combining the ladder diagrams with the random phase
approximation faces the trouble mentioned above. Though our present
investigation is limited to the consideration of the many-boson
systems, the derived result gives a hint that the similar situation
is likely to take place in the Fermi case, too. In this connection
it is worth noting the problem associated with the lack of
self-consistency of the standard method of treating the dilute Fermi
gas~\cite{Fetter}.

The strong-coupling regime is characterized by significant spatial
correlations. So, Eq.~(\ref{10}) resulting in Eq. (\ref{11}) is not
relevant for an arbitrary strong repulsion between bosons at small
separations when we have $\psi(0)=-1, \quad \psi_{{\bf p}}(0)=
-\sqrt{2}$ (see Refs.~\cite{Chern1,Chern2}).
Therefore, to investigate the strong-coupling regime,
Eq.~(\ref{11}) should be abandoned in favor of Eq. (\ref{3}).
Expression (\ref{3}) is accurate to the next-to-leading order in
$(n-n_0)/n$. So, using Eqs. (\ref{3}) and (\ref{8}), we can write
\begin{equation}
g(r)=\varphi^2(r)+\frac{2}{n}\int\frac{d^3q}{(2\pi)^3} n_q
\left(\varphi^2_{{\bf q }/2}({\bf r})-\varphi^2(r) \right).
\label{25}
\end{equation}
Let us now perturb $\widetilde{\psi}(k)$ and $n(k)$. Working to
the first order in the perturbation and keeping in mind conditions
(\ref{12}) and (\ref{25}), from Eq. (\ref{7}) we derive:
\begin{equation}
-2\,T_k \widetilde{\psi}(k)=
    \widetilde{U}(k)(1+2n_k)+2n\,\widetilde{\psi}(k)
                               \widetilde{U}^{\prime}(k)
\label{26}
\end{equation}
with
\begin{equation}
\widetilde{U}(k)=\int \varphi(r) \Phi(r) \exp(-i{\bf k}{\bf r})
d^3r
\label{27}
\end{equation}
and
\begin{eqnarray}
\widetilde{U}^{\prime}(k)=
 \int\Bigl(\varphi_{{\bf k}/2}^2({\bf r})
   -\varphi^2(r)\Bigr)\,\Phi(r)\,d^3r.
\label{28}
\end{eqnarray}
Using Eqs.~(\ref{27}), (\ref{28}) as well
as the relation $\psi_{\bf k}({\bf r}) \to \sqrt{2}\psi(r)\;(k \to
0)$ (see the boundary conditions (\ref{4}))~\cite{note1}, we obtain
$\widetilde{U}(0)\not=\widetilde{U}^{\prime}(0).$ This implies that
the system of Eqs.~(\ref{12}) and (\ref{26}) is not able to yield
the relation $n_k\propto 1/k\;(k \to 0)$ following from the
``$1/k^2$" theorem of Bogoliubov for the zero
temperature~\cite{Bog3}. Indeed, let us assume $n_k \to \infty$ for
$k \to 0.$ Then, from Eq.~(\ref{12}) at $n=n_0$ we find
$n|\widetilde{\psi}(k)|/n_k \to 1$ when $k \to 0.$ On the contrary,
Eq.~(\ref{26}) gives $n|\widetilde{\psi}(k)|/n_k \to
\widetilde{U}(0)/\widetilde{U}^{\prime}(0)\not=1$ for $k \to 0.$
So, consideration of the Bose gas based on Eqs.~(\ref{3}) and
(\ref{12}) does not produce satisfactory results. Nevertheless, it
is worth noting that Eq.~(\ref{26}) has an important peculiarity
which differentiate it from Eq.~(\ref{19}) in an advantageous way.
The point is that in both the limits $n\to 0$ and $k \to \infty$
Eq.~(\ref{26}) is reduced to
\begin{equation}
-\frac{\hbar^2}{m}\,\nabla^2\,\varphi(r)+\Phi(r)\varphi(r)=0.
\label{28aa}
\end{equation}
As it is seen, this is the exact ``bare" (not
in-medium) Schr\"odinger equation, other than its
Born approximation following from Eq. (\ref{20}). Thus, we can expect
the line of our investigation to be right.

As it was shown in the previous paragraph, an approach adequate for
a dilute Bose gas with an arbitrary strong interaction can not be
constructed without modifications of Eq.~(\ref{12}). This is also in
agreement with a consequence of the relation
\begin{equation}
|\langle a_{{\bf k}}\,a_{-{\bf k}}\rangle|^2 \leq
\langle a_{{\bf k}}\,a_{{\bf k}}^{\dagger}\rangle
\langle a_{-{\bf k}}^{\dagger}\,a_{-{\bf k}}\rangle
\label{29}
\end{equation}
resulting from the inequality of
Cauchy-Schwarz-Bo\-go\-liu\-bov~\cite{Bog3}
$$
|\langle \widehat{A}\widehat{B}\rangle|^{2} \leq
    \langle\widehat{A}\widehat{A}^{\dagger}\rangle
    \langle\widehat{B}^{\dagger}\widehat{B}\rangle.
$$
With Eqs. (\ref{6}) and (\ref{29}) one can easily derive $n_0^2
\widetilde{\psi}^2(k) \leq n_k(n_k+1)$. Thus, it is reasonable
to assume that Eq.~(\ref{12}) takes into account only the
condensate-condensate channel and ignores the
supracondensate-condensate ones. Now the question arises
how to find corrections to the r.h.s. of Eq.~(\ref{12}).
At present we have no regular procedure allowing us to do this
in any order of $(n-n_0)/n$. However, there exists an argument
which makes it possible to realize the first step in this
direction. The matter is that the alterations needed have to
produce the equation for $\widetilde{\psi}_{\bf p}({\bf k})$
which is reduced to the equation for $\widetilde{\psi}(k)$
in the limit $p \to 0.$ Though this requirement does not
uniquely determine the corrections to Eq.~(\ref{12}), it
turns out to be significantly restrictive. In particular,
even the simplest variant of correcting Eq.~(\ref{12}) in this
way, leads to promising results. Indeed, this variant is specified
by the expression
\begin{eqnarray}
n_k(n_k+1)= n_0^2\,\widetilde{\psi}^2(k)
+2 n_0\int \frac{d^3q}{(2\pi)^3}\,n_q
                   \widetilde{\psi}^2_{{\bf q}/2}({\bf k}).
\label{30}
\end{eqnarray}
Eq.~(\ref{30}) is valid to the next-to-leading order in
$(n-n_0)/n$. So, we may rewrite it as
\begin{equation}
n_k(n_k+1)=n^2\widetilde{\psi}^2\!(k)+2n\!\int\!
\frac{d^3q}{(2\pi)^3} n_q\!\left(\!\widetilde{\psi}^2_{{\bf q }/2}
({\bf k})-\widetilde{\psi}^2(k)\!\right)\!.
\label{31}
\end{equation}
Perturbing $\widetilde{\psi}(k)$ and $n_k$ and bearing in mind
conditions (\ref{25}) and (\ref{31}), Eq.~(\ref{7}) gives
Eq.~(\ref{26}) again. However, now $\widetilde{U}^{\prime}(k)$
obeys the new relation
\begin{eqnarray}
\widetilde{U}^{\prime}(k)&=&
 \int\Bigl(\varphi_{{\bf k}/2}^2({\bf r})
         -\varphi^2(r)\Bigr)\,\Phi(r)\,d^3r \nonumber\\
&&-\int\frac{d^3q}{(2\pi)^3}
  \frac{\widetilde{U}(q)
    \bigl(\widetilde{\psi}^2_{{\bf k}/2}({\bf q})-
       \widetilde{\psi}^2(q)\bigr)}{\widetilde{\psi}(q)}
\label{33}
\end{eqnarray}
which significantly differs from Eq. (\ref{28}). Indeed, the choice
of the pair wave functions as real quantities implies that operating
with integrands in Eqs. (\ref{27}) and (\ref{33}), one can exploit
$\psi_{{\bf p}}({\bf r})-\sqrt{2} \psi(r) \propto p^2$ at small
$p$~\cite{note3}. For $k \to 0$ this provides
$\widetilde{U}^{\prime}(k)-\widetilde{U}(k)= t_k= c\,k^4 +\cdots$.
Note that for $k \to \infty$ we have $t_k \to -\widetilde{U}(0)$.
Similar to Eq.~(\ref{19}), Eq.~(\ref{26}) can yields results correct
only to the leading order in $(n-n_0)/n$. So, it has to be solved
together with Eq. (\ref{12}) where $n_0^2$ should be replaced by
$n^2$, rather than with Eq. (\ref{31}). This leads to the following
relations:
\begin{eqnarray}
&&n_k=\frac{1}{2}\left(\frac{\widetilde{T}_k+n\widetilde{U}(k)}
{\sqrt{\widetilde{T}_k^2+2n\widetilde{T}_k\widetilde{U}(k)}}
-1\right),\label{35a}\\
&&\widetilde{\psi}(k)=-\frac{
\widetilde{U}(k)}{2\sqrt{\widetilde{T}_k^2+2n\widetilde{T}_k
\widetilde{U}(k)}},
\label{36}
\end{eqnarray}
where $\widetilde{T}_k=T_k+nt_k$, with the limit $\widetilde{T}_k/T_k
\to 1$ at $k \to \infty$.  For $k\to0$ Eq.~(\ref{35a}) gives $n_k
\simeq (\sqrt{n\,m\,\widetilde{U}(0)}/ \hbar k-1)/2$, which is fully
consistent with the ``$1/k^2$" theorem of Bogoliubov for the zero
temperature~\cite{Bog3}.

As it is seen, the strong-coupling regime is more complicated
than the Bogoliubov one because we do not know the quantity
$\widetilde{U}(k)$ {\it ab initio}. To find it, one should solve
Eqs. (\ref{27}) and (\ref{36}) in a self-consistent manner.
Equations (\ref{27}) and (\ref{36}) lead to one more interesting
relation
\begin{equation}
\widetilde{U}(k)=\widetilde{\Phi}(k)-
\frac{1}{2}\int\frac{d^3q}{(2\pi)^3}
          \frac{\widetilde{\Phi}(|{\bf k}
       -{\bf q}|)\widetilde{U}(q)}{\sqrt{\widetilde{T}^2_q+
            2n\widetilde{T}_q\widetilde{U}(q)}}
\label{37}
\end{equation}
which can be called the in-medium Lippmann-Schwinger equation
for the scattering amplitude. To obtain the expansion for the
energy at low densities, we must solve Eq.~(\ref{37}) at $n \to 0$.
Let us rewrite it in the form
$$
\widetilde{U}(k)=\widetilde{\Phi}(k)-
       \frac{1}{2}\int \frac{d^3q}{(2\pi)^3}
           \frac{\widetilde{\Phi}(|{\bf k}
                -{\bf q}|)\widetilde{U}(q)}{T_q}-I_1,
$$
where for $I_1$ we have
$$
I_1=\frac{1}{2}\!\int\!\frac{d^3q}{(2\pi)^3}\!\left\{
    \frac{\widetilde{\Phi}(|{\bf k}
       -{\bf q}|)\widetilde{U}(q)}{\sqrt{\widetilde{T}^2_q+
            2n\widetilde{T}_q\widetilde{U}(q)}}-
              \frac{\widetilde{\Phi}
             (|{\bf k}-{\bf q}|)\widetilde{U}(q)}{T_q}\!\right\}.
$$
Operating with $I_1$ in the same manner as we dealt with $I$
in the section~\ref{3sec} and taking into account that $t_k=0$ at
$k=0$, for $n \to 0$ we derive
\begin{equation}
I_1= - \alpha \widetilde{\Phi}(k), \quad
\alpha=\frac{\sqrt{n m^3}}{\pi^2\hbar^3}\widetilde{U}^{3/2}(0).
\label{38}
\end{equation}
From Eqs.~(\ref{37}) and (\ref{38}) it now follows that
\begin{eqnarray}
\widetilde{U}(k)-\widetilde{U}^{(0)}(k)&=&
\alpha\widetilde{\Phi}(k)              -\int \frac{d^3q}{(2\pi)^3}
\frac{\widetilde{\Phi}(|{\bf k}
   -{\bf q}|)}{2T_q}\nonumber \\
&&\times\Bigl(\widetilde{U}(q)
             -\widetilde{U}^{(0)}(q)\Bigr).
\label{39}
\end{eqnarray}
Here $\widetilde{U}^{(0)}(k)=\int \varphi^{(0)}(r) \Phi(r)
\exp(-i{\bf k}{\bf r})d^3r$ but now $\varphi^{(0)}(r)$ obeys
Eq.~(\ref{28aa}) rather than Eq.~(\ref{20}) taken in the limit
$n \to 0$ like in the section~\ref{3sec}. Let us introduce
the new quantity $\widetilde{\xi}(q)=-(\widetilde{U}(q)-
\widetilde{U}^{(0)}(q))/2T_q$. Then, for its Fourier transform
$\xi(r)$ we obtain
\begin{equation}
-\frac{\hbar^2}{m}\nabla^2\Bigl(\alpha + \xi(r)\Bigr)
                    +\Phi(r)\Bigl(\alpha+\xi(r)\Bigr)=0,
\label{40}
\end{equation}
here $\xi(r) \to 0$ when $r \to \infty$. Comparing
Eq.~(\ref{40}) with Eq. (\ref{28aa}), we find $\xi(r)=\alpha
\psi^{(0)}(r)$. Hence, for $n \to 0$ we have
\begin{equation}
\widetilde{U}(k) \simeq \widetilde{U}^{(0)}(k)\left(1
+\gamma(k,n) \frac{8}{\sqrt{\pi}}\sqrt{na^3}\right).
\label{41}
\end{equation}
Here $\gamma(k,n) \to 1$ when $n \to 0$. Besides, the relation
$\widetilde{U}^{(0)}(0)=4\pi \hbar^2 a/m$ is used in
Eq.~(\ref{41}), where $a$ is the scattering length.

Having in our disposal Eq.~(\ref{41}), we are able to calculate
the expansion in powers of $n$ for the condensate depletion and
energy of a dilute Bose gas with an arbitrary strong interparticle
potential. Considering the condensate depletion $(n-n_0)/n=
1/(2\pi)^3\int_{0}^{+\infty}dk\, 4\pi k^2 n_k/n$, with the help of
Eq. (\ref{35a}) we obtain
\begin{equation}
\frac{n-n_0}{n}=\frac{8}{3\sqrt{\pi}}\sqrt{n a^3}+\cdots.
\label{42}
\end{equation}
Notice that according to Eq. (\ref{41}) one can expect that among
the omitted terms in Eq.~(\ref{42}) there is one proportional to
$n a^3$.

The most simple way of deriving the expansion for the
mean energy per particle is based on using the chemical potential
which, in the presence of the Bose condensate, is given by
\begin{equation}
\mu=\frac{1}{\sqrt{n_0}}\int d^3 r^{\prime}
   \Phi(|{\bf r}-{\bf r}^{\prime}|)
      \langle \psi^{\dagger}({\bf r}^{\prime})
           \psi({\bf r}^{\prime}) \psi({\bf r})\rangle.
\label{43}\end{equation}
This formula follows from the well-known expression
$
\delta \Omega = \langle\delta
                 \left(\hat H-\mu \hat N\right) \rangle,
$
where $\delta\Omega$ is an infinitesimal change of the grand
canonical potential, and relation~(see Ref. \cite{Bog3})
$$
\frac{\partial \Omega(N_0,\mu,T)}{\partial N_0}=0.
$$
Using the specific expressions for the scattering parts of the
condensate-condensate and supracondensate-condensate pair
wave functions \cite{Chern1} given by Eqs. (\ref{6}) and (\ref{44})
one can represent Eq.~(\ref{43}) in the following form:
\begin{equation}
\mu=n_0 \widetilde{U}(0)+
    \sqrt{2}\int \frac{d^3q}{(2\pi)^3}\;n_q\;
         \widetilde{U}_{{\bf q}/2}({\bf q}/2),
\label{45}\end{equation}
here
$$
\widetilde{U}_{{\bf p}}({\bf k})=
\int \varphi_{\bf p}({\bf r}) \Phi(r) \exp(-i{\bf k}{\bf r}) d^3 r.
$$
Now, for $n \to 0$ (see the procedure of calculating the integral $I$
in the section~\ref{2sec} one can rewrite Eq.~(\ref{45}) as
\begin{equation}
\mu=n\widetilde{U}(0)\left(1+\frac{n-n_0}{n}+ \cdots\right).
\label{46}\end{equation}
Inserting Eqs. (\ref{41}) and (\ref{42}) into Eq. (\ref{46}),
we arrive at
\begin{equation}
\mu=\frac{4\pi \hbar^2 a n}{m}\left(1+\frac{32}{3\sqrt{\pi}}
\sqrt{n a^3}+\cdots\right).
\label{47}
\end{equation}
This result for the chemical potential implies, due to the
basic thermodynamic formula $\mu=\partial(\varepsilon n)/\partial n$,
the following expansion for the mean energy per particle:
\begin{equation}
\varepsilon=\frac{2 \pi \hbar^2 a n}{m}\left(1+
       \frac{128}{15\sqrt{\pi}}\sqrt{n a^3}+\cdots\right).
\label{48}\end{equation}

As it is seen, the relation (\ref{48}) coincides with the well-known
result of the approach~\cite{Lee} being reduced to the Bogoliubov
model with the ``dressed'' interaction. It is not a surprise because
according to the conclusions of the section~\ref{2sec}, we know that
the numerical factor $128/(15\sqrt{\pi})$ appears in the series
expansion for the mean energy per particle within the Bogoliubov
model (see Eq. (\ref{21i})).  Replacing the bare interaction
potential by the ``dressed'' one results in replacing the scattering
length $a_B$ in Eq.(\ref{21i}) by its exact value $a$. The only
problem of doing so concerns the parameter $b$. Indeed, it follows
from Eq. (\ref{21h}) that substituting the hard-sphere potential
$\widetilde{U}(0)=4\pi\hbar^2 a/m$ for $\widetilde{\Phi}(k)$ leads
to the familiar divergency~(see, e.g. Ref. \cite{Fetter}, p. 314).
This obstacle has been overcome with the help of the well-known
argument of Landau~(see the footnote in the paper~\cite{Bog1})
stating that the thermodynamics of dilute quantum gases is only
ruled by the vacuum scattering amplitude. According to this
reasoning one can expect that dependence on the shape of the
interaction potential should not appear in the first orders of the
density series expansion of the thermodynamic quantities. So,
various regularizing procedures, more or less speculative, have been
worked out in order to exclude this divergence~(together with the
parameter $b$). On the contrary, there are no problems like this
within the approach of the present paper.  Here Eq.~(\ref{48}) is
derived on the solid theoretical basis rather than with the help of
Landau's argument. In spite of its reasonable character, it needed
to be corroborated, and the results of this paper given by
Eqs.~(\ref{42}), (\ref{46}) and (\ref{48}) have proved the validity
of Landau's argument beyond any inconsistencies and divergencies. In
the weak-coupling case when $|\Phi(r)| \ll 1$, the energy per
particle calculated within our scheme is expressed by Eq.~(\ref{48})
with $a$ replaced by $a_B$.  So, the appearance of the parameter $b$
in the results of the section \ref{2sec} is an artifact following
from the neglect of scattering in the supracondensate-condensate
pair wave channel.

The divergence mentioned in the previous paragraph is not typical of
the strong-coupling perturbation theory for the many-boson systems
but results from, say, the weak-coupling spirit of the approach of
Ref. \cite{Lee}. A simple way to be convinced of this is to consider
the spatial boson correlations. Taken to the lowest-order with
respect to the density, the structural factor (see the last paper in
Ref.~\cite{Lee}) is of the form
\begin{equation}
S(k)=\frac{T_k}{\sqrt{T_k^2+2nT_k \widetilde{U}^{(0)}(k)}}.
\label{49}
\end{equation}
By definition we have
\begin{equation}
g(r)=1+\frac{1}{n} \int \frac{d^3k}{(2\pi)^3} (S(k)-1)
                                       \exp(i{\bf k}{\bf r}).
\label{50}
\end{equation}
Using Eqs. (\ref{49}) and (\ref{50}), for $n \to 0$ one can readily
find
\begin{equation}
g(r) \to 1+2\psi^{(0)}(r),
\label{51}
\end{equation}
where $\psi^{(0)}(r)$ obeys Eq.~(\ref{28aa}). This result answers
the approximation (\ref{11}) while $\psi^{(0)}(r)$ is not related to
the weak-coupling regime and obeys the exact ``bare'' Schr\"odinger
equation. In the situation $\Phi(r) \to \infty$ for $r \to 0$ one
has $\psi^{(0)}(r=0)=-1$, which implies, according to Eq.
(\ref{51}), $g(r=0) \to -1$ for $n \to 0.$ It is not consistent with
the physical sense of $g(r)$ and has nothing to do with the
strong-coupling case corresponding to Eq. (\ref{25}) when for
$n \to 0$
$$
g(r) \to \left(1+\psi^{(0)}(r)\right)^2.
$$
Notice that the zero-density limits for the thermodynamic quantities
of a strongly interacting dilute Bose gas were first found in the
Bogoliubov original paper~\cite{Bog1}:
$$
(n-n_0)/n\to 0,\ g(r)\to(\varphi^{(0)}(r))^{2},\
\varepsilon/n\to \widetilde{U}^{(0)}(0)/2.
$$
At last, we remark that due to the incorrect picture of the spatial
boson correlations found in papers~\cite{Lee}, one can expect
significant alterations for the spectrum of the elementary
excitations too. However, to clarify these corrections we should
conclusively solve the problem concerning relation between the
momentum distribution and scattering parts of the pair wave
functions. Indeed, it has been mentioned that there exist various
possibilities of generalizing Eq.~(\ref{12}) so as to obtain the
equation for $\widetilde{\psi}_{\bf p}({\bf k})$ being reduced to
the equation for $\widetilde{\psi}(k)$ in the limit $p \to 0.$ These
possibilities result in the same series expansions for the
thermodynamic quantities (\ref{42}), (\ref{47}) and (\ref{48}) but
produce different data for the long-range spatial boson
correlations. Here we limited ourselves to considering the most
simple variant of generalizing Eq. (\ref{12}), which makes it
possible to investigate only the thermodynamics of a strongly
interacting Bose gas. The interesting and important problem of the
spectrum of the elementary excitations is thus beyond the scope of
this paper and will be the subject of the future investigations.
%-----------------------------------------------------------------

\section{Conclusion}
\label{5sec}

Concluding let us take notice of the important points of this paper
once more. It was demonstrated that thermodynamically consistent
calculations based on Eqs. (\ref{11}) and (\ref{12}) conclusively
result in Eqs.~(\ref{19})-(\ref{21}).  Therefore, using the
Bogoliubov model with the ``dressed" interaction does not provide
the satisfactory solution of the problem of the strong-coupling Bose
gas. As it was shown, when investigating this subject, one should go
beyond the Bogoliubov scheme. To do this, we developed the approach
reduced to the system of Eqs.~(\ref{27}), (\ref{33}), (\ref{35a})
and (\ref{36}). These equations leading to the in-medium
Lippmann-Schwinger equation (\ref{37}), reproduce the familiar
results (\ref{42}), (\ref{46}) and (\ref{48}) for the condensate
depletion, chemical potential and mean energy but yield completely
different picture of the spatial boson correlations. This difference
should manifest itself in the next orders of the density series
expansions for the thermodynamic quantities and in the excitation
spectrum as well.

This work was supported by the RFBR Grant No. 97-02-16705.

%-------------------------------------------------------------------

\end{multicols}

\begin{thebibliography}{99}
\bibitem{note4} By the leading order we mean zero depletion
of the condensate: $n=n_{0}$.
\bibitem{note3} It is not true in the case of complex-value
wave functions. However, in this situation integrands in Eqs.
(\ref{27}) and (\ref{33}) would include the absolute values
of $\widetilde{\psi}(k)$ and $\widetilde{\psi}_{{\bf p}}({\bf k})$
for which we have $|\widetilde{\psi}_{{\bf p}}({\bf k})|-\sqrt{2}
|\widetilde{\psi}(k)| \propto p^2$ at small $p$ in the case of the
boundary conditions (\ref{4}).
\end{thebibliography}
\end{document}